\documentclass{aa}
\usepackage{graphicx, epsfig,psfig}
\usepackage{subfigure}

\usepackage{natbib}
\usepackage{references}
\bibpunct{(}{)}{;}{a}{}{,} % to follow the A&A style  

\newcommand{\lineA}[2]{\mbox{\,#1\,{\sc #2}}}

\begin{document}

   \title{Observations and modelling of a large optical flare on AT Microscopii}

   \author{D. Garc\'{\i}a-Alvarez
          \inst{1}
          \and
          D. Jevremovi\'c\inst{1}\inst{,}\inst{2}
	  \and J.G. Doyle\inst{1}
	  \and C.J. Butler\inst{1}}        

   \offprints{D. Garc\'{\i}a-Alvarez}

   \institute{Armagh Observatory, 
              College Hill, Armagh BT61 9DG N.Ireland\\
	      \email{dga@star.arm.ac.uk, djc@star.arm.ac.uk, jgd@star.arm.ac.uk, cjb@star.arm.ac.uk}
         \and
             Astronomical Observatory, Volgina 7, 11070 Belgrade, Yugoslavia}

   \date{Received ....., 2001; accepted ...., .....}

   \abstract{
Spectroscopic observations covering the wavelength range 3600--4600\AA\ are presented for a large flare on the late type M dwarf AT Mic (dM4.5e). A procedure to estimate the physical parameters of the
flaring plasma has been used which assumes a simplified slab model of the flare based on a comparison of observed and computed Balmer decrements. With
this procedure we have determined the electron density, electron temperature, optical thickness and temperature of the
underlying source for the impulsive and gradual phases of the  flare. 
The magnitude and duration of the flare allows
us to trace the physical parameters of the response of the lower atmosphere.  In order to check our
derived values we have compared them with other methods. In addition, we have also applied our procedure to a stellar and a solar
flare for which parameters have been obtained
using other techniques. 
   }

   \maketitle

\section{Introduction}

Stellar flares are events where a large amount of energy is released in a 
short interval of time, radiating at almost all frequencies in the 
electromagnetic spectrum. Flares are believed to result from the release of 
magnetic energy stored in the corona through reconnection
\citep[see reviews by][]{Mirzoyan84,Butler91,Haisch91,Garcia-Alvarez00}.
Many types of cool stars produce flares \citep{Pettersen89}, sometimes at 
levels several orders of magnitude more energetic than their solar counterparts. 
The exact mechanism(s) leading to the energy release and subsequent excitation 
of various emission features remains poorly understood. In the dMe stars (or 
UV Cet type stars) optical flares are a common phenomenon. In more luminous 
stars, flares are usually only detected through UV or X-ray observations \citep{Doyle89b}, 
although optical flares have been detected in young early 
K dwarfs like LQ Hya \citep{Montes99}.

One would like to trace all the energetic processes in a flare to some common 
origin likely to be localized at some unresolvably fine scale in the magnetic 
field. {The largest solar flares observed involve energies of  
$\sim10^{32}\ \rm{erg}$ \citep{Gershberg89}, while large flares on dMe stars can be one  
order of magnitude larger \citep{Doyle90,Byrne90}}. Even more energetic flares occur on the RS CVn 
binary systems, the total flare energy in the largest of this type of systems 
may exceed E$\sim 10^{38}\ \rm{erg}$ \citep{Doyle92,Foing94}. Such a change 
in the star's radiation field modifies drastically the 
atmospheric properties over large areas, from photospheric to coronal layers. 
{Models by \citet{Houdebine92} indicate that heating may propagate down to low 
photospheric levels, with densities higher than $10^{16}\ \rm{cm}^{-3}$,
although electrons with energies in the $\rm{MeV}$ range would be required to 
attain such depth.} 

As regards the derivation of physical parameters such as electron density and
temperature, various methods have {been} used. For example, \citet{Katsova90}
published an analysis of flare Balmer decrements. The broadening and merging 
of higher Balmer lines, dominated by the Stark effect, have also {been} used 
to estimate electron densities in the chromosphere \citep{Donati-Falchi85}. 
Measurements of the broadening of the lower Balmer lines,
which are less affected by the Stark effect, together with line
shifts, provide information on the large scale motions during flares. {Steep 
decrements are evidence for electron densities between
$10^{8}$ and $10^{12}\ \rm{cm}^{-3}$, while a shallow Balmer decrement 
indicates densities larger than $10^{13}\ \rm{cm}^{-3}$ in the flaring M dwarf 
atmosphere.}

\citet{Jevremovic98} and \citet{Jevremovic99} developed a
procedure to fit the Balmer decrement  
based on the solution of the radiative transfer equation using the escape 
probability computing technique of \citet{Drake80} and \citet{DrakeUlrich80} 
and the direct search method of \citet{Torczon91,Torczon92}. 

In this paper we present the results of a large flare observed on AT Mic 
(Gliese 799B, V=10.25, B=11.83 $ 20^h41^m51^s, -32^o26'07'' $, Equinox 2000). 
It is a visual binary star with a separation of 4.0 arcsec \citet{Wilson78},
approximately 10.2 pc away, with both components of spectral type dM4.5e and 
subject to flaring \citep{Joy49}. \citet{Kunkel70} determined the flare 
incidence at 2.8 flares per hour. From its position in the (Mv, R-I) diagram, 
\citet{Kunkel73} deduced that AT Mic lies slightly above the
main sequence, pointing to its probable membership of the young disc 
population. \citet{Nelson86} reported 29 U-band/B-band flares in  39.3
hours, only one of which showed microwave flaring at 5 \rm{GHz}. They also 
reported two microwave bursts without any optical counterpart. \citet{Kundu87} 
found that both components of the AT Mic system were active and 
variable at 6 and 20 $\rm{cm}$ wavelengths. The peak emission levels
were detected in the southern component (Gliese 799B). X-ray and ultraviolet 
emission from this system has been observed several times over the
last ten years or so \citep{Linsky82,Pallavicini90}. 

In $\S$2 we present the {current} dataset plus a discussion of the data analysis 
while $\S$3 gives the results. In $\S$4, we analyse the evolution of the main physical 
parameters during the flare using the code developed by 
\citet{Jevremovic98} and apply the procedure to a stellar and 
a solar flare for which parameters have been obtained using other techniques. The
conclusions are given in $\S$5.   

\section{Observations and Data Reduction}
The observations were carried out at the South African Astronomical Observatory 
between 20--26 August 1985. Optical spectroscopy was obtained with the 1.9m 
telescope at SAAO equipped with the RGO Unit Spectrograph and the Reticon Photon
Counting System (RPCS), providing spectral coverage from 3600 to 4600\AA\ at 
a reciprocal dispersion of 50\AA $/\rm{mm}$ and a wavelength resolution of 
$\Delta \lambda \sim$ 2\AA\ (FWHM) at H$\gamma$. The spectral region includes 
the entire Balmer series except H$\alpha$ and H$\beta$, as well as the 
\lineA{Ca}{ii} H \& K lines and the \lineA{He}{i} 4026\AA\ and 
\lineA{He}{i} 4471\AA\ lines. The exposure time was 60 sec. which with
readout overheads gave a time resolution of 70 sec. During the run on AT Mic, sky conditions were generally 
good although seeing was variable.

Spectra were reduced in the standard manner, consisting of  background
subtraction and flat-field correction using exposures of a tungsten lamp.
Wavelength calibration was made with Cu--Ar lamp spectra taken immediately
preceeding and following the stellar spectra. {A first-order spline cubic 
fit to some 16 lines achieved a nominal wavelength calibration accuracy of
0.12 \AA. An atmospheric extinction 
correction for SAAO was also applied to the data. Flux 
calibration, by reference to spectrophotometric standard
stars (LTT7987 and LTT9239), 
was carried out using the NOAO packages IRAF.\footnote{IRAF is distributed by the National 
Optical Astronomy Observatories,
    which are operated by the Association of Universities for Research
    in Astronomy, Inc., under cooperative agreement with the National
    Science Foundation.} 
The observations were made in spectrophotometric mode with a slit of 400$\mu$ 
width ($\sim\,4\ \rm{arc\,sec}$), adequate to encompass the entire image during the worst seeing conditions 
encountered during the run. A dekker was selected which provided an approximately square aperture. The programme star (AT Mic) 
and the spectrophotometric standards were observed under similar conditions. From a comparison of the calibrations afforded 
by the two standards, we estimate the spectrophotometric accuracy to vary from $\pm\, 0.12$ magnitudes ($\sim\,12\%$) at 3600\AA\ 
to $\pm\,$0.03 magnitudes ($\sim\,3\%$) at 4600\AA. The fluxes of individual emission lines will have additional errors associated with 
the photon noise level of the binned fluxes.}

%=======================  FIG. 1 ================
\begin{figure}[t!]
   \centering
   \resizebox{\hsize}{!}{\includegraphics{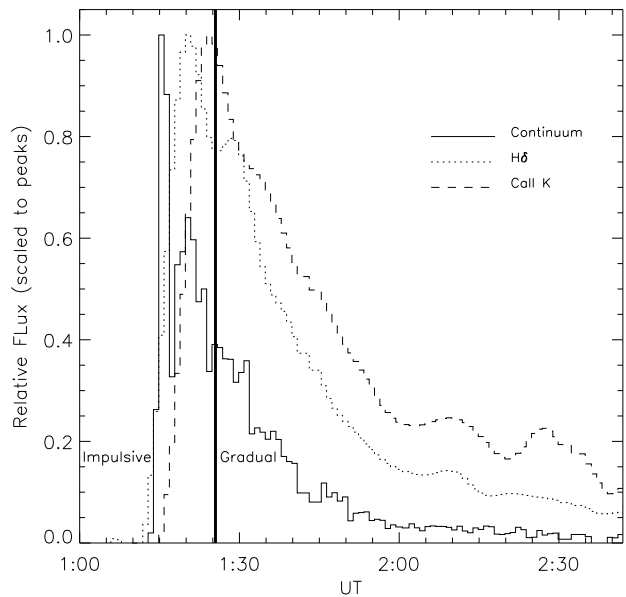}}
   \caption{Normalised {flare-only} light curves for H$\delta$, \lineA{Ca}{ii} K
   and the continuum (3600-3700\AA). The demarcation between the impulsive and 
   gradual phases, taking into account the different line behaviour, is indicated by a vertical line. }
              \label{FigImpGrad}%
    \end{figure}

\section{AT Mic flare at 1:20 UT on 21 August 1985}

{In this paper we report on an energetic flare, lasting more than one and a half hours, observed on 21 August 1985. 
The continuum emission, which initially peaked in the impulsive phase at 1:15 UT, had a second peak at 1:20 UT, roughly 
coincident with the primary peak in the Balmer lines. In addition, there is evidence for a second peak at 1:30 UT in some of the 
Balmer lines (H$\delta$, H8) and also in \lineA{He}{i} 4026\AA.} Fig.~\ref{FigImpGrad} shows a rapid continuum 
flux rise and decay, while the hydrogen Balmer lines peak later
and decay more slowly. There is some indication that the low-order Balmer lines
decay more slowly than the high-order lines, as has been noted by \citet{Butler91} in a flare on YZ CMi. The \lineA{Ca}{ii} K line rises 
and decays even more slower than the hydrogen lines. Following \citet{Hawley91}, we 
define the start of the gradual phase as the time when the continuum 
shows a turnover from fast to slow decay. The change from impulsive to gradual 
phase is indicated in Fig.~\ref{FigImpGrad} with a solid vertical line.

\begin{figure*}[th!]
   \centering
   \includegraphics[width=17cm]{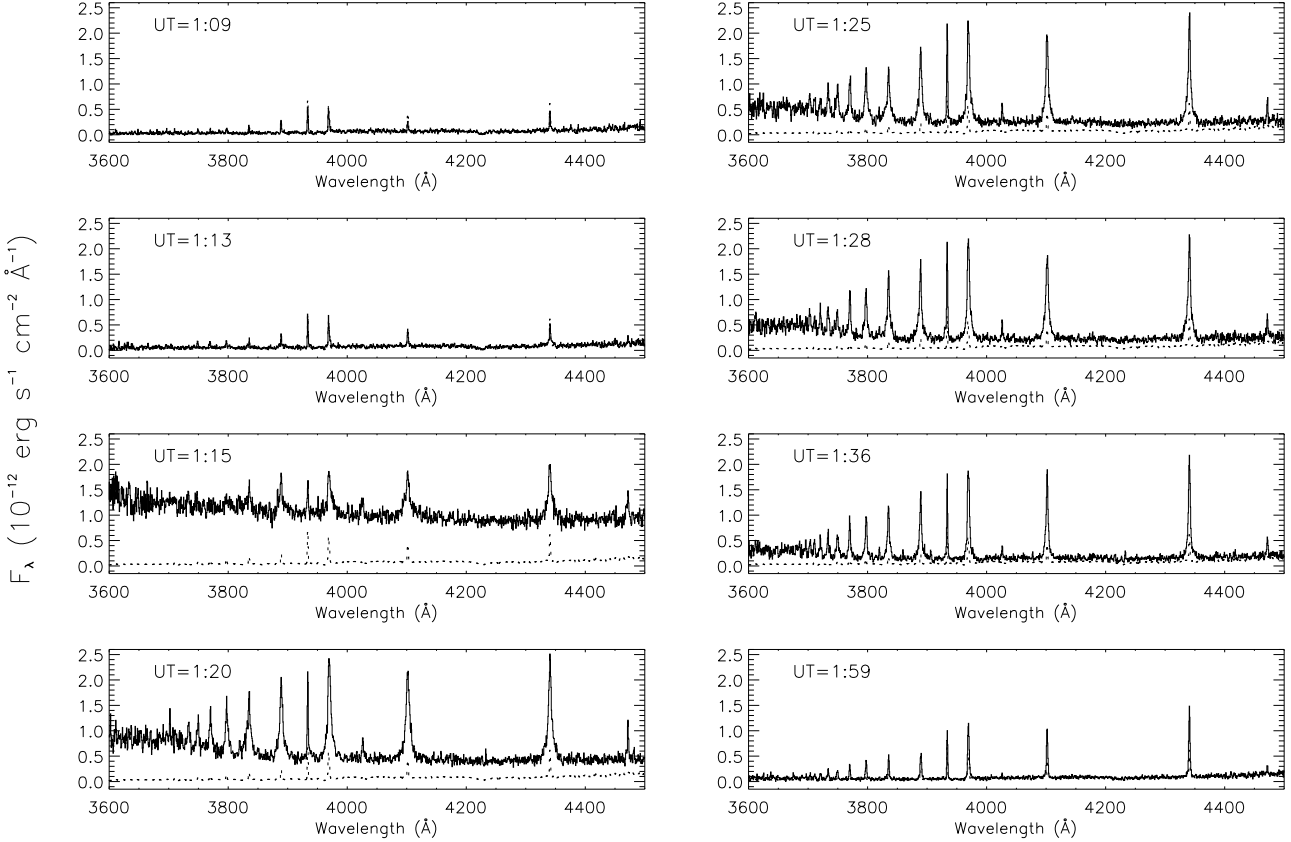}
   \caption{The time sequence of optical {flare-only} spectra of AT Mic  
during the pre-flare phase, impulsive 
phase, maximum continuum, maximum flux in H$\delta$, minimum between the two
maxima, second maximum flux in H$\delta$, the beginning of the gradual phase and 
finally the gradual phase. The mean quiescent spectrum is shown as a dotted line 
for comparison.
}
              \label{FigFlareSequence}%
    \end{figure*}

Many emission lines other than the Balmer line series and \lineA{Ca}{ii} 
H \& K were seen during this flare, e.g. \lineA{Fe}{i} and \lineA{He}{i}. Many 
of these were visible only during the impulsive phase and at the beginning of 
the gradual phase. Observed at higher spectral resolution these lines would 
provide important information on the changing atmospheric structure
and hence flare heating. Fig.~\ref{FigFlareSequence} illustrates a 
time sequence of the optical {flare-only} spectra obtained during the AT Mic flare. 
%==============================================================================
%=======================  SEVERAL SPECTRUM DURING FLARE  =========================

%

\begin{table*}[th!]
\begin{center}
	\caption[]{The total energy radiated in the Balmer, Calcium and Helium 
	lines during the flare on AT Mic. Units are 10$^{30}\ \rm{erg}$}
\begin{tabular}{ccccccccccc} 
\hline  \hline \\
 UT    & H$\gamma$ & H$\delta$ & H8  & H9  & H10 & \lineA{Ca}{ii} K & \lineA{Ca}{ii} H+H$\epsilon$ & \lineA{Ca}{i} & \lineA{He}{i}
 4026$\, \AA$ & \lineA{He}{i} 4471$\, \AA$ \\

\hline \\
1:13 &  1.61 &	1.42 &      .74 &	.46 &	   .24 &  .04 &	       .36 & .04 &       .52 &       .90  \\
1:15 &  4.62 &	4.37 &     2.50&       2.10&	   .87 &  .25 &       3.76 & .01 &      1.34&	    1.43   \\
1:20 & 11.07&       10.69&	   7.54&       5.28&	  2.83 & 2.28&       9.86  & .36 &       .67 &      1.24  \\
1:25 &  9.03 &	8.25 &     6.12&       5.24&	  3.23 & 2.42&       8.28  & .14 &       .56 &       .94    \\
1:28 &  8.12 &	8.52 &     6.32&       4.91&	  2.84 & 2.60&       8.10  & .41 &       .61 &       .90    \\
1:30 &  6.30 &	5.22 &     4.05&       3.42&	  2.08 & 1.63&       5.94  & .34 &       .48 &       .48    \\
2:00 &  1.81 &	1.55 &      .97 &	.85 &	   .65 &  .51 &       1.85 & .11 &       .13 &       .39    \\

\hline
\end{tabular}
\end{center}

\end{table*}

\begin{figure*}[th!]
   \centering
   \includegraphics[width=17cm]{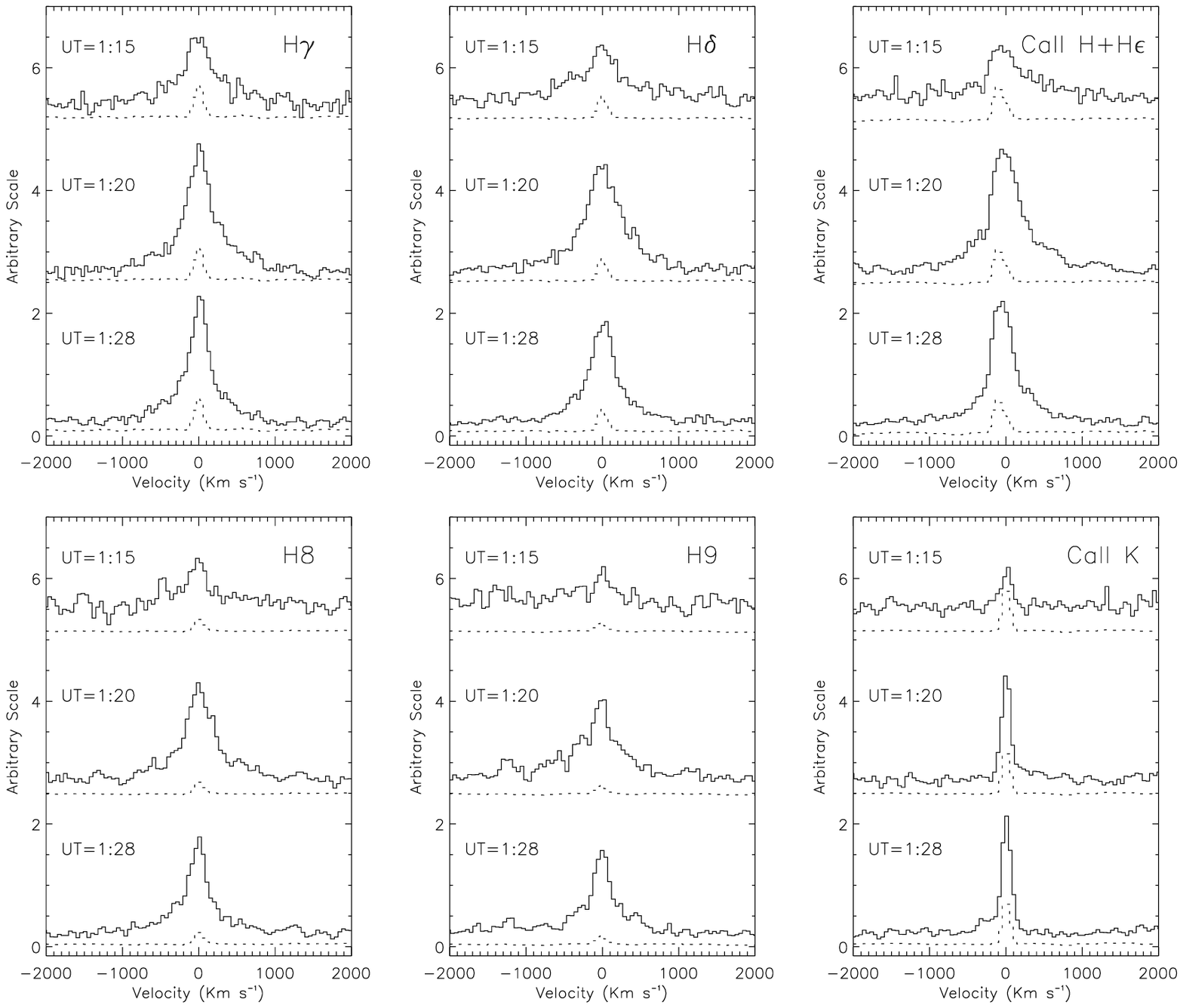}
   \caption{Comparison of the mean quiescent and {flare-only} line profile 
   for the Balmer series lines, Ca\,{\sc ii} H\,+\,H$\epsilon$ and Ca\,{\sc ii} K.}
 
              \label{FigLineCenter}%
    \end{figure*}
In order to form a mean quiescent spectrum we have co-added 75 quiescent 
spectra. The Balmer lines from H$\gamma$ to H13 and the 
\lineA{Ca}{ii} H \& K lines are visible and in emission in the quiescent 
spectrum. In the flare spectrum the Balmer lines increase their emission and 
there is a substantial enhancement of the ultraviolet continuum. {There is also 
some evidence for a filling in of the strong absorption 
line \lineA{Ca}{i} at 4227\AA\ observed during the impulsive and gradual 
phase of the flare}. The 
\lineA{He}{i} 4026\AA\ and \lineA{He}{i} 4471\AA\ lines are also in 
emission during the flare. 

Emission line fluxes were measured as a function of time for the important 
contributors to the radiative losses in the optical, including the  hydrogen 
Balmer lines from H$\gamma$ to H9, the \lineA{Ca}{ii} H \& K lines and the
Helium lines (see Table~1). To analyze the optical spectra, the mean quiescent spectrum was 
first subtracted from the spectra taken during the flare. The emission line 
fluxes were then determined by numerically integrating the flux under the line, 
using carefully chosen windows to set the local continuum level.  
Our low resolution spectroscopy cannot resolve 
the \lineA{Ca}{ii} H and H$\epsilon$ line, but the blend is dominated
 by the behavior of H$\epsilon$. We note the comparatively slow response 
of the \lineA{Ca}{ii} K line compared to the Balmer lines.

Both the Balmer lines and \lineA{Ca}{ii} emission lines 
were above the quiescent values for more than 2 hours, while the continuum 
enhancement lasted around one hour.

\subsection{Line Broadening}

Near flare maximum, the wings of H$\alpha$, H$\delta$, H8 and even H9 are 
very broad, around 25\AA\ at the base, {as is shown in Fig.~\ref{FigLineCenter}}. 
As a result, the Balmer lines start  
to merge at about H13; meanwhile \lineA{Ca} {ii} K
shows very little broadening. \citet{Giampapa83} and \citet{Worden83} pointed 
out that during flares the lower Balmer lines such as H$\alpha$ and H$\beta$ were 
unbroadened, while the higher Balmer series lines show significant broadening, 20\AA\ 
or more for some lines. They also reported that the Balmer lines (and 
\lineA{He}{i}) in a very large flare were broadened and asymmetric to the red, 
while \lineA{Ca}{ii} is unbroadened. Balmer line broadening in solar flares 
is generally attributed to Stark broadening \citep{Svestka72}. 

\begin{figure}[t!]
   \centering
   \resizebox{\hsize}{!}{\includegraphics{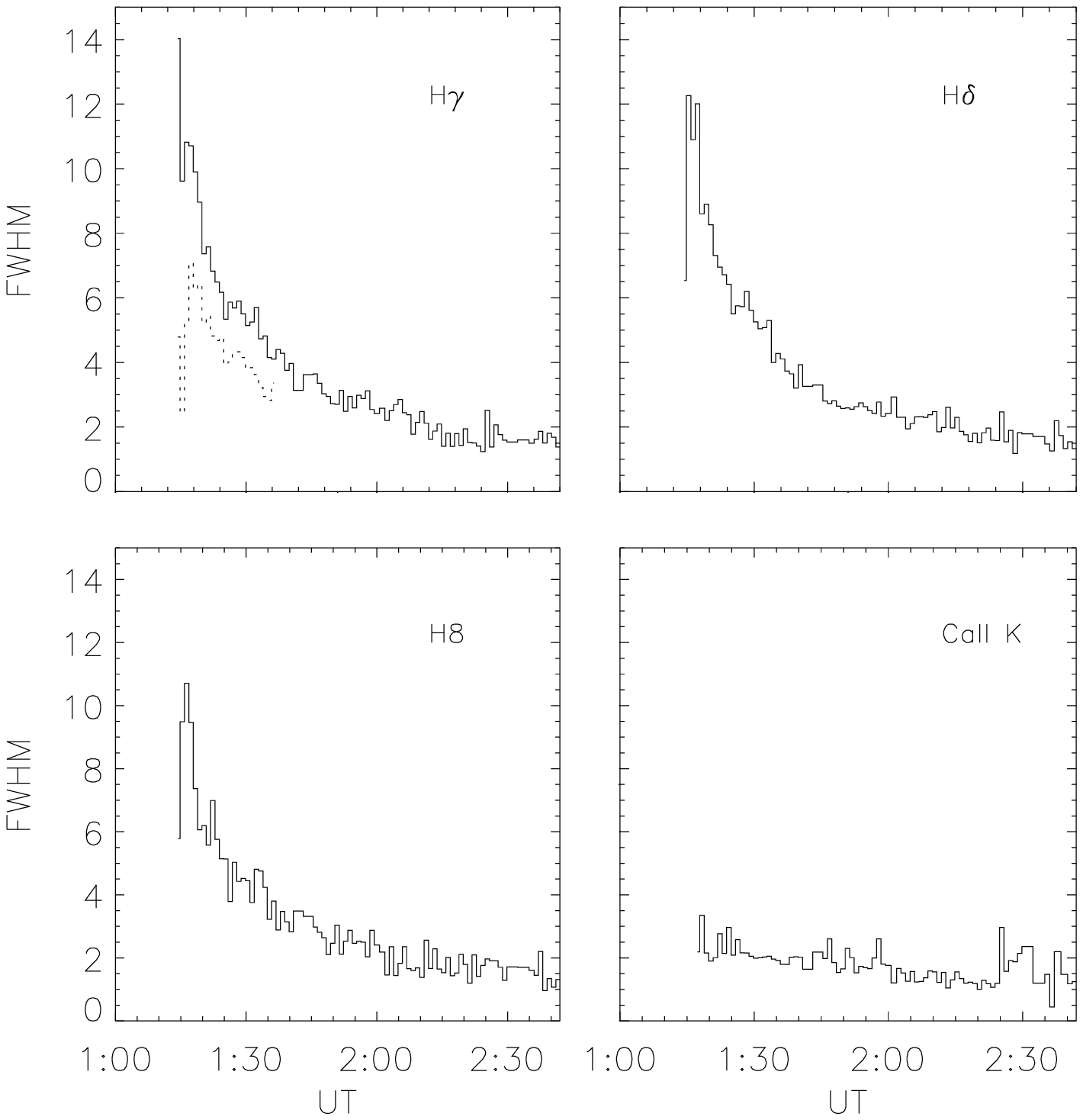}}
   \caption{FWHM through the flare-only spectrum on AT Mic for H$\gamma$, 
   H$\delta$, H8 and \lineA{Ca} {\sc ii} K lines.
   For the Balmer lines we have used a Lorentzian profile and Gaussian profile 
   for the Ca {\sc ii} K line. In the top left we plot the Voigt profile
   fitting (dotted) for H$\gamma$.}
              \label{FigFWHM}%
    \end{figure}

In order to obtain an estimate of the broadening of the hydrogen lines during the flare, we 
have measured the full width at half maximum (FWHM) for H$\gamma$, H$\delta$, H8 and 
\lineA{Ca}{ii} K. The FWHM for the Balmer lines were obtained using a 
Lorentzian profile, while a Gaussian profile was used for \lineA{Ca}{ii} K. 
From Fig.~\ref{FigFWHM} we can conclude that the FWHM of the emission lines 
increases dramatically during the flare for the Balmer lines, although for \lineA{Ca}{ii} K 
there is almost no change.

It seems that line broadening continues well into the gradual phase of the 
flare, finishing around the same time as the continuum enhancement. As \citet{Doyle88} 
pointed out, a possible explanation for the line wing enhancement could be 
motions of several hundred $\rm{km\, s^{-1}}$. If that were the case in the 
AT Mic flare, it would require the amount of radiating material moving upwards 
and downwards to be approximately equal throughout the flare in order to produce 
the observed profiles. Although this could be the case it has not been observed on 
the Sun, where normally a pronounced redshift in H$\alpha$ is usually observed 
during the impulsive phase \citep{Zarro89,Wuelser89}.

If hydrodynamic effects were contributing significantly in the chromospheric 
line forming regions, we would expect \lineA{Ca}{ii} K to show similar broadening 
to the Balmer lines, but this was not observed. {Although the 
broadening of the Balmer series lines decreases almost exponentially during and after the 
impulsive phase, the central line intensity of H$\delta$ (Fig.~\ref{FigImpGrad}) shows 
two peaks, indicating that there are two energy 
releases. The first burst induced line broadening while the second burst did not, suggesting 
that this flare may have derived from an arcade of loops.}

The excess {broadening} of the Balmer lines {could} be interpreted as turbulence 
as we did not observe any clear asymmetry at any time, in any line, during the 
flare. 

{Our conclusions in the last two sections, regarding the evolution and broadening of the emission lines 
during the flare on AT Mic, are similar to those of \citet{Hawley91} for a flare on AD Leo.}
 
%==============================================================================
%=======================  FWHM FOR SEVERAL LINES ==============================
%
\subsection{The \lineA{He}{i} 4026\AA\ and \lineA{He}{i} 4471\AA\ Triplet Lines}
Helium lines are 
produced under generally higher excitation conditions than other lines in the 
visible and the lines are generally  optically thin, which makes interpretation 
easier. {The helium lines are also useful as indicators of the presence of 
electric fields in the flaring plasma due to their forbidden components. However, even for the Sun, exact knowledge of the
background spectrum is required for the determination of the contribution of 
the forbidden lines.} The \lineA{He}{i} lines have been reported in solar white light flares
\citep{Lites86} and in stellar flares \citep{Moffet76,Doyle88}. \citet{Lites86} pointed out that any mechanism which produces a 
dense plasma at relatively high temperatures ($2\ 10^{4}-10^{5}\ \rm{K}$) will produce emission in
neutral helium.
\begin{figure}[t!]
   \centering
     \resizebox{\hsize}{!}{\includegraphics{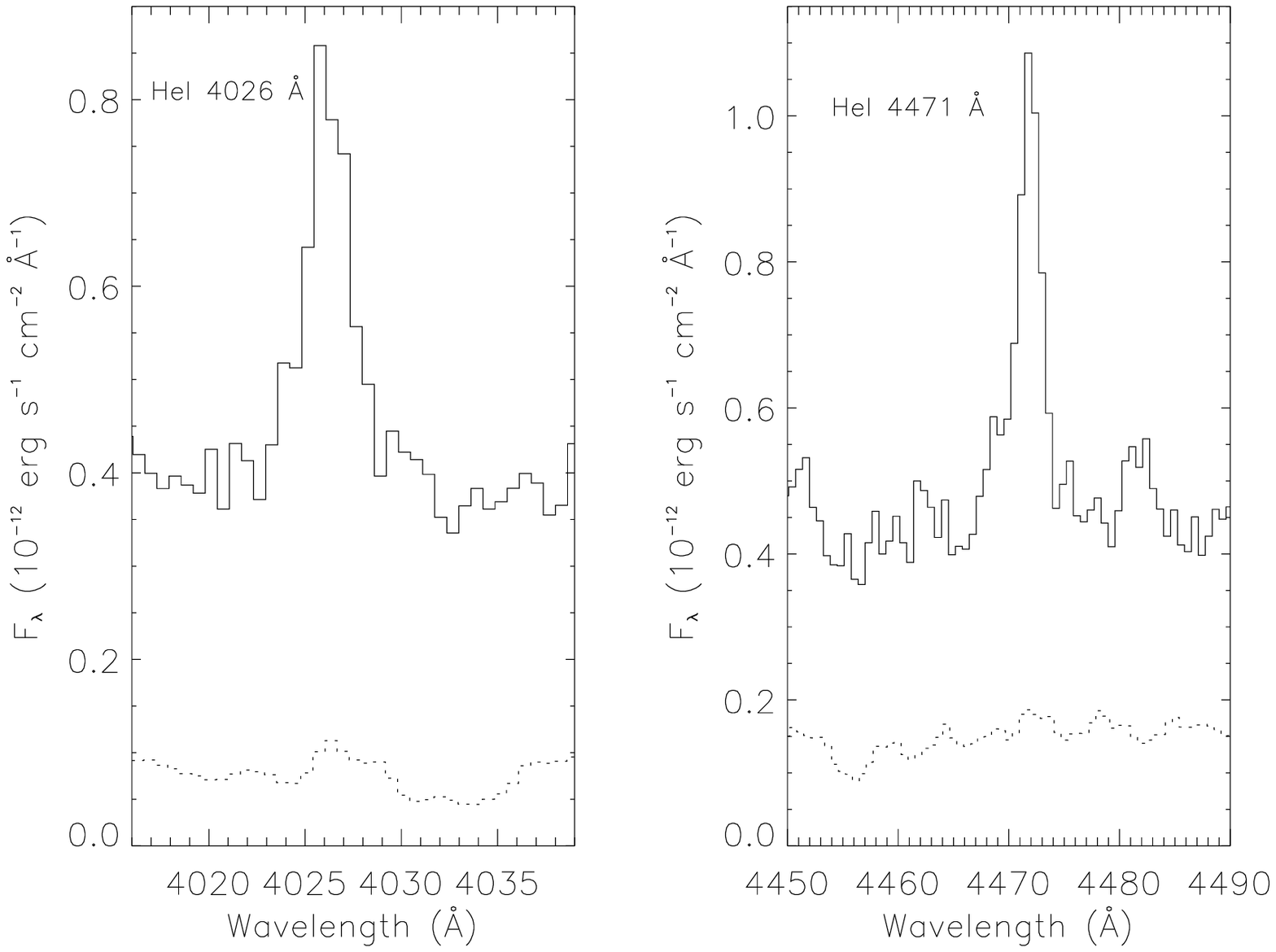}}  
   \caption{Left panel: Average over 
   4 flare-only spectra of \lineA{He}{i} 4026\AA\ around flare maximum. 
   Right panel: \lineA{He}{i} 4471\AA\ average over 4 flare-only spectra around 
   flare maximum.}
 
              \label{FigCaihei}%
    \end{figure}

The \lineA{He}{i} 4026\AA\ line is a diffuse triplet between levels 
$2^3$P and $5^3$D. This line is  
usually not detected except in the strongest events. {It was shown \citep[e.g.][]{Vidal64} 
that the higher members of this triplet 
are quite sensitive to density changes. The \lineA{He}{i} 
4026\AA\ line is near to a strong \lineA{Mn}{i} absorption feature at 
4030\AA\ (which during intense activity can go 
into emission due to pumping from \lineA{Mg}{ii} k, \citep{Doyle92,Doyle01}) 
thus making detection difficult. We observed the \lineA{He}{i} 4026\AA\ line and 
\lineA{He}{i} 4471\AA\ line in emission during the impulsive phase and well into the
gradual phase of the flare on AT Mic (Fig.~\ref{FigCaihei}). Two peaks are seen in the time profile of 
\lineA{He}{i} 4026\AA, the first at 1:20 UT and the second at 1:30 UT. These are simultaneous with the maxima in the 
Balmer lines. In \lineA{He}{i} 4471\AA\, only a single peak at 1:20 UT is visible.} 

{The FWHM of the two \lineA{He}{i} lines 
are 5.0\AA\ and 2.8\AA\ respectively. \citet{Gieske69} give a procedure for computing the line profile of \lineA{He}{i} 4026\AA\ 
including Stark and Doppler broadening. With their model for $T_{e} = 20\,000\ \rm{K}$, we derive an upper limit for the 
electron density of $10^{16}\ \rm{cm^{-3}}$ for the \lineA{He}{i} emitting region.}

\subsection{Continuum Analysis}
\begin{figure}[t]
   \centering
   \resizebox{\hsize}{!}{\includegraphics{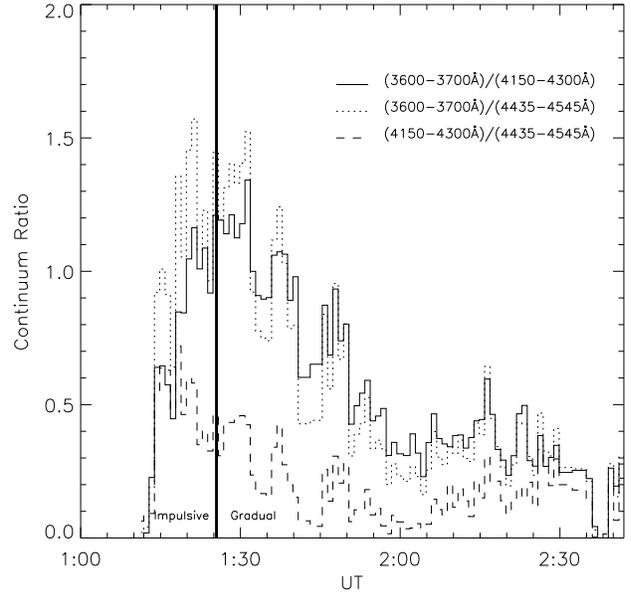}}
   \caption{Continuum flux ratios through the flare on AT Mic. It is clear from the plot that the continuum of the flare 
   has a sustantial blue component during both the impulsive phase and gradual phase and that the maximum blueness in the near UV occurs
   close to the end of the impulsive phase.}
              \label{FigThreeContinuum}%
    \end{figure}

{We have measured the fluxes (Table~2) for three different continuum regions: one in the near ultraviolet, 3600--3700\AA\, 
and two in the blue, 4130-4300\AA\ (with the exclusion of the 
\lineA{Ca}{i} line region) and 4435-4545\AA. Fig.~\ref{FigThreeContinuum} 
shows the variation in the ratios of the continuum fluxes during the impulsive and gradual phases of the flare. 
We note that the ratios of the near ultraviolet band to the two longer wavelength bands do not reach their maximum blueness 
until the end of the impulsive phase and the beggining ot the gradual phase (indicated by the vertical line), whereas the ratio 
of the intermediate wavelength band to the longer wavelength band reaches it maximum blueness at the beginning of the impulsive phase. 
As all three continuum bands reach their maximum brightness at the start of the impulsive phase, this behaviour suggest that a simple 
heating of a static, confined, kernel with radiation following a black-body law, is not appropiate \citep[see][]{Butler96}.} 

{\citet{Bopp73} 
suggested that the continuum enhancement ends when the fast decay 
(what we term the impulsive phase) has ended. In fact, this generally accepted 
notion, that the white-light continuum is present only during the impulsive 
phase is certainly not the case for the AT Mic flare nor for the flare on AD Leo reported by \citet{Hawley91}.} 

Spectroscopic observations of M dwarf flares reveal the shape of the flare 
continuum and allow us to explore the possible continuous emission mechanisms. 
{It has been suggested \citep[e.g.,][]{CramWoods82,Houdebine92} that the optical flare continuum is 
composed of bound-free and free-free emission from a two-component thermal 
model characterized by electron temperatures of $T\sim 10^{5}\ \rm{K}$ and 
$T\,\sim\,2\,10^{4}\ \rm{K}$. \citet{CramWoods82} showed the spectral energy 
distributions of some of the suggested radiative processes that may give rise 
to the flare continuum. The shape of the free-free curve at $T\sim 10^{5}\ 
\rm{K}$  obtained by \citet{Giampapa83} produces reasonable fits to the 
observed flare continuum. \citet{Kunkel70} has argued that hydrogen free-bound 
emission is the principal component for the observed continuum emission in M 
dwarf flare events. The curve for this process given by \citet{Giampapa83} only 
fits the red part of the continuum, which would imply that additional continuum
processes, other than bound-free, are taking place in higher temperature 
layers. \citet{Hawley92} fitted the observed flare continuum, but the results were much bluer than those computed from the models. 
They concluded that the best fit to their observations was a
blackbody spectrum with $T\,\sim\,8500-9500\ \rm{K}$. \citet{Hawley92} proposed that the flare continuum is formed
by photospheric reprocessing of intense UV/EUV line emission from the upper chromosphere.} 

%==============================================================================
%=======================  THREE CONTINUUM  ============================

%
%
%==============================================================================

\begin{table}[t!]
\begin{center}
	\caption[]{Continuum energy radiated during quiescence and during 
	flare on AT Mic. Units are 10$^{32}\ \rm{erg}$}
\begin{tabular}{cccc} 
\hline  \hline \\
 UT    & 3600--3700\AA\ & 4150--4300\AA\ & 4435--4545\AA \\

\hline \\
Quiescence & .03 &	   .05 &	 .12   \\
1:13     & .04 &	   .05 &	 .11   \\
1:15     & .98 &	   .68 &	 .74   \\
1:20     & .63 &	   .31 &	 .35   \\
1:25     & .40 &	   .17 &	 .22   \\
1:28     & .37 &	   .16 &	 .21   \\
1:36     & .23 &	   .11 &	 .16   \\
2:00     & .06 &	   .06 &         .12   \\

\hline
\end{tabular}
\end{center}

\end{table}

\citet{Dame85} proposed a model for solar white-light flares in order to explain the observed blue continuum. Using the study of
radiative processes in solar flares \citep{Dame83} they show how existing semiempirical models can be  modified to predict a visible spectrum more
similar to the WLFs emission observed. The hot temperature plateau (6--9$\,10^{4}\ \rm{K})$ model proposed by \citet{Dame85} is not valid
for all WLFs but provides an explanation for the origin of an unidentified source of opacity distinct from the Balmer and Paschen
continuum. {\citet{Houdebine96} suggested that this blackbody type emission is due to the enhanced 
electron density in the lower chrosmophere.}

However, the blue continuum enhancement could also be due to merging of the 
higher members of the Balmer lines particularly in the vicinity of the
Balmer jump, which indicates that these lines contribute a non-negligible amount 
to the near-ultraviolet continuum {\citep{Donati-Falchi85,Zarro85}}. One case has been reported in which the continuum was found to be red in
color during the first stages of the flare \citep{Doyle89a}.

\begin{table*}[th!]
\label{Tabcodecomparation}
\begin{center}
	\caption[]{Comparison of physical parameters for flares on YZ CMi and 
	on the Sun obtained with the different codes.}
\begin{tabular}{lcccccc} 
\hline  \hline \\
 Star    & Code applied & Reference &  $N_e$ (\rm{cm$^{-3}$}) &  $T_e (\rm{K})$  & $\tau(Ly\alpha)$ & Area(\rm{cm$^{2}$}) \\ 

\hline \\
YZ CMi  & Our Code         & 1 & 5\,10$^{13}$     &  13\,000  & 3\,10$^{6}$ & 1\,10$^{18}$ \\
        & Gas Dynamic      & 2 &    10$^{14}$     &  10\,000  & 2\,10$^{6}$ & 5\,10$^{17}$ \\ \\
The Sun & Our Code         & 1 & 2\,10$^{14}$     &  18\,000  & 7\,10$^{4}$ & 7\,10$^{19}$ \\ 
        & Stark Broadening & 3 & (2-3)\,10$^{13}$ &  14\,000  & n/a         & 5\,10$^{19}$ \\ 

\hline
\end{tabular}
\end{center}
\hspace{2.3cm} {\it References}: 1: This work, 2: \citet{Katsova91}, 3: \citet{Johns-Krull97}.\\

\end{table*}

\section{Derivation of Flare Parameters}

We normalized the flare-only fluxes for H$\delta$, H8, H9 and H10 to 
H$\gamma$ at several times during the impulsive and gradual phases. We do not 
include H11 and higher Balmer lines due to the difficulty in assigning the 
local continuum level. The Balmer decrement became quite shallow and stayed 
nearly constant during the impulsive phase of the flare and well into the 
gradual phase. After this, the decrement steepens but does not become as 
steep as that during the quiescent phase.

\subsection{Our Procedure}
\citet{Jevremovic98} developed a procedure to fit the Balmer decrements. It is based 
on the solution of the radiative transfer equation, using the escape 
probability technique of \citet{Drake80}, and \citet{DrakeUlrich80}. 
The \citet{DrakeUlrich80} 
method assumes:(i) that the escape probability for a slab of 
optical thickness is averaged over all directions with a uniform distribution of 
emitters, (ii) complete redistribution of frequencies and {(iii) that the line profile 
is represented by a Doppler core and power law wings of slope $-$5/2
to emulate the Stark effect.} 
\citet{Jevremovic98} adopted a simplified picture of the flaring plasma as a 
slab of hydrogen with an underlying thermal source of radiation and which causes 
photoionization. {This source, which can be considered to be the photosphere, can 
be exposed to additional heating during the flare. It assumes that the Balmer lines are
formed in the upper  
chromosphere. We note that, this simplified picture leaves out some
important processes (i.e. cooling from Mg\,{\sc ii}, Ca\,{\sc ii} and
other metals). However, as seen from Table~1 the AT Mic flare is
predominantly seen in Balmer lines and we believe that this picture may be applicable.}
The procedure minimizes the difference between the observed and 
calculated Balmer decrement using a multi-directional search algorithm 
\citep{Torczon91,Torczon92}. This allows us to find the best possible solution for the 
Balmer decrement in four parameter space where the parameters are: electron 
temperature, electron density, optical depth in the Ly$\alpha$ line and the 
temperature of the underlying source. Details of the procedure and the 
multi-search algorithm and comparison with the classical simplex algorithm are 
given in \citet{Jevremovic98}. From the best solutions for the Balmer 
decrement, the procedure allows us to calculate the effective thickness of the
slabs of hydrogen plasma with which we are able to determine the 
surface area of the emitting plasma from the calculated emission measure. 
{Although the temperature of the underlying source could be a free parameter
in our code, we have fixed a lower limit at $5\,500\ \rm{K}$ for the Sun and $2\,500\ 
\rm{K}$ for YZ CMi and AT Mic, for numerical stability reasons.}

\subsection{Comparison with other Codes}
In order to check the \citet{Jevremovic98} procedure we have
applied it to two flares for which physical parameters have been
independently derived, one stellar and one solar.

\subsubsection{Comparison with the Gas-Dynamic Model for the YZ CMi flare}

We have applied our procedure to a flare on YZ CMi observed during a 
coordinated run involving telescopes at the South African Astronomical 
Observatory, and the ESA X-ray satellite, EXOSAT, on 4 March 1985. The photometric 
observations consisted of UBVRI photometry with a 0.75-m
telescope. Details have  been published by \citet{Doyle88}. The
spectra of the flare on YZ CMi, over  
the wavelength range 3600--4400\AA\, were analysed by \citet{Katsova91} 
using a gas-dynamic model. Based on the gas-dynamic model 
and using the data set for the flare observed by \citet{Doyle88} on YZ CMi, 
\citet{Katsova91} derived the main physical parameters at flare maximum. 
They matched quite well the observed data with the theoretical decrement for a 
temperature $T=10^{4}\ \rm{K}$, electron density $10^{14}\ 
\rm{cm^{-3}}$ and optical depth of the layer at the $Ly\alpha$ line center 
$\tau_{Ly\alpha}=2\,10^{6}$. They also derived the emitting source area of 
$>5$x$10^{17}\ \rm{cm^{2}}$ using a description of the optical continuum as 
a Planck function for a temperature of $10^{4}\ \rm{K}$. 

Our values for the derived flare parameters are reasonably close to those  values obtained by \citet{Katsova91} 
(see Table~3), although our values for the electron 
temperature and $Ly\alpha$ optical depth at line center are slightly larger 
and smaller respectively. We obtained an effective thickness of the slab of 
hydrogen plasma of around $2\,000$ $\rm{km}$ and an emissivity of $3.25\ 
\rm{erg}\,\rm{cm}^{-3}\,\rm{s}^{-1}$. 

\subsubsection{Comparison with a Stark Broadening Model for a Solar flare}

As a second check we have also applied our procedure to a solar flare observed 
on 6 March 1993. Details of the flare have been published by \citet{Johns-Krull97}. They fitted the observations with models 
that include Stark broadening by electrons, assuming that the Balmer lines form in a homogeneous flaring region overlying the quiet
photosphere that has $n_{e}\approx (2-5)\,10^{13}\ \rm{cm}^{-3}$. They 
modelled the line widths following the general procedure outlined by \citet{Redman54} and \citet{Svestka63}, 
which attempts to model the flare emission profile by taking into account the effects of 
self-absorption, Doppler broadening and Stark broadening, and comparing the resulting profiles with the observed flare-quiet Sun profiles. 
\citet{Johns-Krull97} found that, over the course of the first 10 
flare spectra, the electron density varied fairly randomly in the range $(2-3)\,10^{13}\ \rm{cm}^{-3}$. Meanwhile, the values for the
electron temperature were around $14\,000\ \rm{K}$. Although they did not calculate the 
flare area they provided an H$\alpha$ image taken at BBSO around 10 minutes after the flare maximum and from this image one can infer a
solar flare area of about $5\,10^{19}\ \rm{cm}^{2}$. 

{Using our procedure, and the solar flare data obtained by \citet{Johns-Krull97}, we obtain the parameters at 
solar flare maximum as shown in Table~3. As in the 
case of the YZ CMi flare, our values are close to those obtained by \citet{Johns-Krull97} although 
our value for the electron density is almost one order of magnitude larger. This result for the electron density could be due 
to the fact that, in our fit of the solar flare, we have not fitted the
lower Balmer lines, H$\alpha$ and H$\beta$ which are formed higher in the atmosphere, and were observed by \citet{Johns-Krull97}.
We have also discarded H7 in our fitting, 
in order to compare the results of both solar and stellar flares based on the same Balmer lines.} The flare area 
is in very good agreement with that derived from the observations. We obtained an  
effective thickness for the hydrogen slab of almost 100
 $\rm{km}$ and an emissivity of 
$145\  \rm{erg}\,\rm{cm}^{-3}\,\rm{s}^{-1}$.

%==============================================================================
%=======================  MODELLING OTHER FLARES = ============================

%
\begin{figure}[t]
   \centering
   \resizebox{\hsize}{!}{\includegraphics{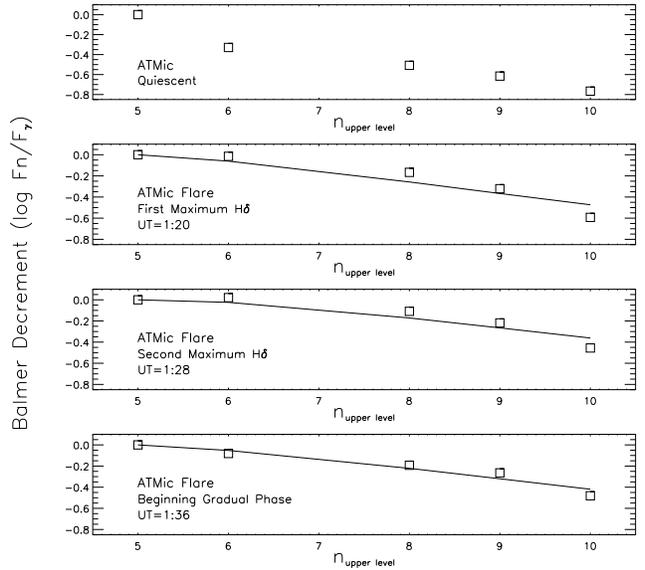}}
   \caption{The observed Balmer decrements (squares) and the optimum computed 
   fits (solid line) during quiescence and at different times during the flare.}
              \label{FigModelatmic}%
    \end{figure}
%
%==============================================================================

\subsection{Modelling the AT Mic flare}

\begin{figure*}[t!]
   \centering
   \includegraphics[width=17cm]{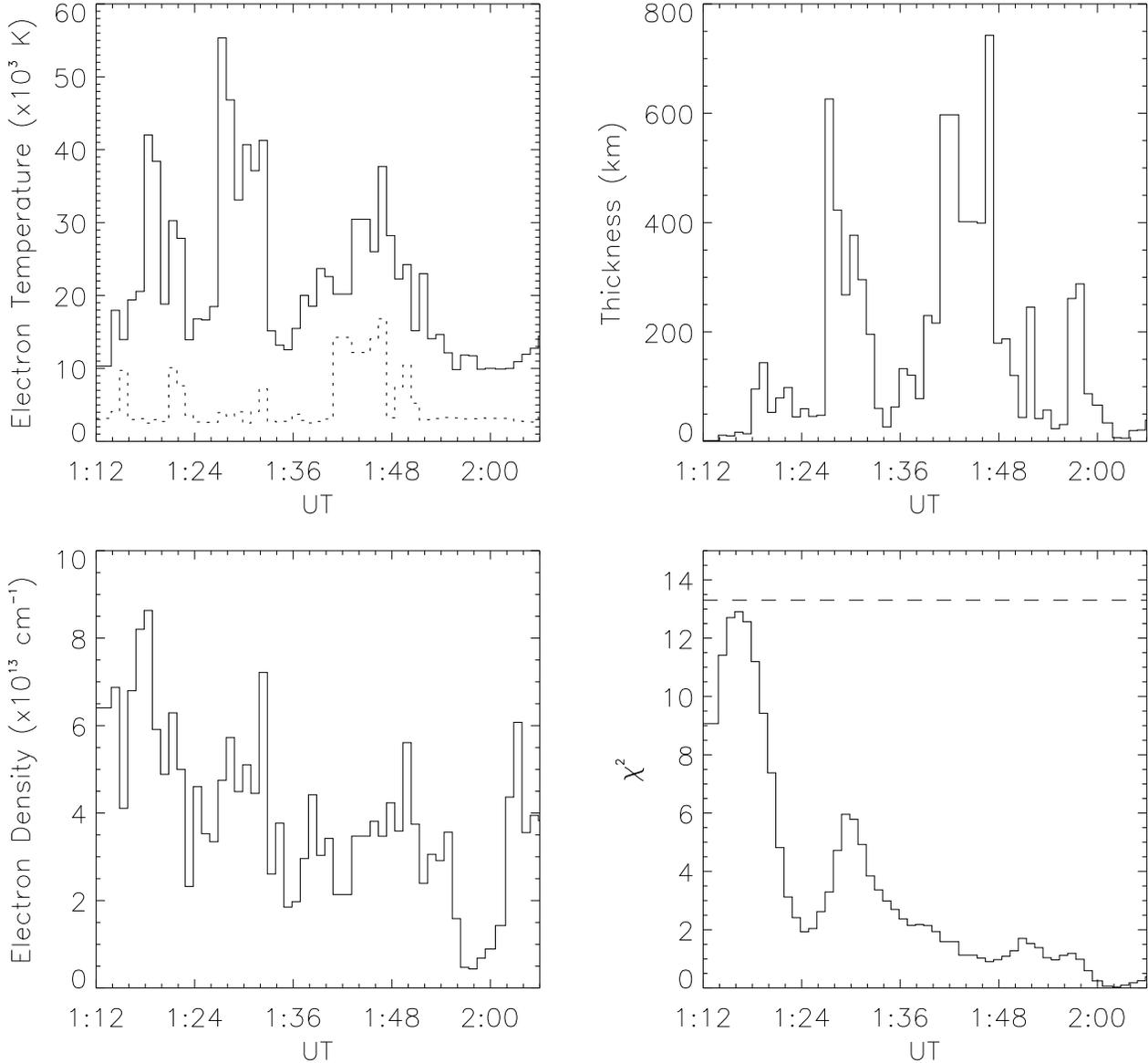}
%\begin{figure*}[t!]
%\psfig{file=/project6/djc/home/papers/atmic/svi3.ps, width=14cm, angle=270}
   \caption{Parameters derived from the fit to the Balmer decrement
using the direct search method. Evolution of the electron temperature
and temperature of background source are shown in the top left panel,
electron density in bottom left panel , effective plasma thickness in
top right panel and goodness of fit in bottom right panel.}
              \label{FigParametersFlare}%
    \end{figure*}

From the above two cases we concluded that the procedure of \citet{Jevremovic98} 
gives reliable results and we proceed to apply it to the AT Mic flare. 
In Fig.~\ref{FigModelatmic} we have plotted the data and three Balmer decrement 
fits during the flare and one during quiescence. The evolution  of the main 
parameters is given in  
Fig.~\ref{FigParametersFlare}. 

Analysing the Balmer decrement during the AT Mic flare, we observed
that, at the  
beginning of the impulsive phase, the Balmer decrement has a steep slope which 
become shallower as the flare progresses through its impulsive phase. The Balmer decrement keeps almost 
the same slope until well into the gradual phase, after which the
slope steepens reaching similar values to that in the quiescent
phase. In the two spectra prior to the first H$\delta$ maximum at 1:20 and in the three 
spectra after the second H$\delta$ maximum at 1:28, the
H$\delta$ flux is larger than the H$\gamma$ flux. {Due to their similarity, we have compared the observed Balmer decrement behaviour 
for the flare on AD Leo reported by \citet{Hawley91} and the observed Balmer decrement for our flare on AT Mic. During the quiescent 
phase, the slope for AT Mic is steeper than for AD Leo, while, during the impulsive and 
gradual phases, the slopes for AT Mic are 
shallower than thoses for AD Leo. This means that the variations of the physical parameters during 
the AT Mic flare are larger than during 
the AD Leo flare.}

%=======================  PARAMETERS AT Mic FLARE= ============================

%==============================================================================

{From Fig.~\ref{FigParametersFlare} one can see that there are three
distinct peaks in the electron 
temperature, of which two coincide with peaks in H$\delta$. 
We interpret these two peaks as the result of two sudden separate releases of
energy (which presumably correspond to the two continuum peaks shown in Fig.~\ref{FigImpGrad}), while the third peak
in the temperature is connected to additional, more gradual
heating. This is also seen in the behaviour of the temperature of
the background source and in Fig. \ref{FigThreeContinuum}. 
Note that peaks in the slab thickness coincide with the
maximum temperature, which can be interpreted by a change in the degree of ionization with temperature. An increase in the degree of
ionization leads to a thicker slab for the same optical depth. Also,
the thickness of the slab, coupled with the low emissivity of the plasma,
explains the relatively low increase in the continuum/line flux during the third peak 
(see Fig. \ref{FigImpGrad}). For example, the modeled emissivity in H$\gamma$ for
the first peak was around $150\ \rm{erg\, cm^{-3}\, s^{-3}}$ while during the
third event the emissivity was around $10\ \rm{erg\, cm^{-3}\, s^{-3}}$. Similar
volumes of plasma for both peaks lead to much lower increase in the
flux during the third event (contrary to what one would expect taking
into account only  thermal radiation).  Also, one has 
to keep in mind that  our background source is included only as the source of 
photoionization which controls ionization balance. Inclusion of other sources
of ionization (i.e. focused beam of electrons/protons) would
probably change the results of modeling, but that is beyond the scope
of this paper. }

We found little change in the electron density during the flare(s) and
it stays between $10^{13}--10^{14}\ \rm{cm^{-3}}$. These values are
one or two orders of magnitude below the upper limits deduced from either
Stark broadening or the Inglis-Teller relation. This confirms the
reliability in our method, based in the simultaneous analysis of a few lines, rather than some earlier estimates of electron density
\citep[i.e.][]{Houdebine92}, based on a single line analysis.

From the best fit solutions for the Balmer decrement, we have calculate the 
effective thickness of the slab of hydrogen plasma from the beginning
of the impulsive phase until the beginning of the gradual phase as shown
in the top right panel of Fig.\ref{FigParametersFlare}. Using these 
results we are able to determine the surface area of the
emitting plasma from the calculated emission measure and the measured
fluxes in H$\gamma$. From these results we conclude 
that the area of the flare reach three maximum values, which take place 
after each maxima in the electron temperature.  The values we obtain are
around few times $10^{19}\ \rm{cm}^{2}$ for all of them. Taking into 
account that the radius of AT Mic is $0.32\ R_{\sun}$, this suggests that the
flares occupy around 0.6 percent of the stellar surface
area. This is comparable with the sizes of the largest 
solar flares \citep{Tandberg-Hanssen88}.

\section{Conclusions \& Discussion}

We observed a major flare on AT Mic in the 3600--4500\AA\ wavelength region 
which allowed us to compute a detailed flare energy budget in the optical 
spectral region. The physical parameters obtained are consistent with previously
derived values for stellar flares. We have compared the results from our 
method for two flares previously analysed, namely, one on YZ CMi and one on 
the Sun, obtaining reasonable agreement. 
 
We have obtained the first detailed trace of physical parameters during a 
stellar flare. We have also shown that our method could provide a suitable 
model for the study of flares on stars of different spectral type (G-M). The total 
radiated energy released in the region 3600-4500\AA\ during the flare was 
$\sim4\,10^{33}\ \rm{erg}$ and $\Delta U\sim\,4\, \rm{mag}$. The 
energy radiated by this flare is comparable with the strongest solar ones but
is of medium strength compared to flares on other dMe stars.

Further stellar observations with high time and high spectral resolution 
should be obtained. Also, further solar observations should be obtained in 
order to check the method with other examples, particularly, the flare area 
and thickness.

\begin{acknowledgements}
      Research at Armagh Observatory is grant-aided by the Department of 
      Culture, Arts and Leisure for N. Ireland. This work was supported in part
      by PPARC grant PPA/G/S/1997/00298. We thank the SAAO for the provision 
      of telescope time and Dr. E.R. Houdebine for useful discussions. {We also thank 
      the referee S.L. Hawley for helpful comments.}
\end{acknowledgements}

\bibliographystyle{apj} % style apj.bst
\bibliography{references}

\begin{thebibliography}{55}
\expandafter\ifx\csname natexlab\endcsname\relax\def\natexlab#1{#1}\fi

\bibitem[{{Bopp} \& {Moffett}(1973)}]{Bopp73}
{Bopp}, B.~W. \& {Moffett}, T.~J. 1973, \apj, 185, 239

\bibitem[{{Butler}(1991)}]{Butler91}
{Butler}, C.~J. 1991, Memorie della Societa Astronomica Italiana, 62, 243

\bibitem[{{Butler}(1996)}]{Butler96}
{Butler}, C.~J. 1996, in IAU Colloq. 153: Magnetodynamic Phenomena in the Solar
  Atmosphere - Prototypes of Stellar Magnetic Activity, 217

\bibitem[{{Byrne} \& {McKay}(1990)}]{Byrne90}
{Byrne}, P.~B. \& {McKay}, D. 1990, \aap, 227, 490

\bibitem[{{Cram} \& {Woods}(1982)}]{CramWoods82}
{Cram}, L.~E. \& {Woods}, D.~T. 1982, \apj, 257, 269

\bibitem[{{Dame} \& {Cram}(1983)}]{Dame83}
{Dame}, L. \& {Cram}, L. 1983, \solphys, 87, 329

\bibitem[{{Dame} \& {Vial}(1985)}]{Dame85}
{Dame}, L. \& {Vial}, J.-C. 1985, \apjl, 299, L103

\bibitem[{{Donati-Falchi} {et~al.}(1985){Donati-Falchi}, {Falciani}, \&
  {Smaldone}}]{Donati-Falchi85}
{Donati-Falchi}, A., {Falciani}, R., \& {Smaldone}, L.~A. 1985, \aap, 152, 165

\bibitem[{{Doyle} {et~al.}(1988){Doyle}, {Butler}, {Bryne}, \& {van den
  Oord}}]{Doyle88}
{Doyle}, J.~G., {Butler}, C.~J., {Bryne}, P.~B., \& {van den Oord}, G.~H.~J.
  1988, \aap, 193, 229

\bibitem[{{Doyle} {et~al.}(1989{\natexlab{a}}){Doyle}, {Butler}, \& {van den
  Oord}}]{Doyle89a}
{Doyle}, J.~G., {Butler}, C.~J., \& {van den Oord}, G.~H.~J.
  1989{\natexlab{a}}, \aap, 208, 208

\bibitem[{{Doyle} {et~al.}(1989{\natexlab{b}}){Doyle}, {Byrne}, \& {van den
  Oord}}]{Doyle89b}
{Doyle}, J.~G., {Byrne}, P.~B., \& {van den Oord}, G.~H.~J. 1989{\natexlab{b}},
  \aap, 224, 153

\bibitem[{{Doyle} {et~al.}(2001){Doyle}, {Jevremovi{\' c}}, {Short},
  {Hauschildt}, {Livingston}, \& {Vince}}]{Doyle01}
{Doyle}, J.~G., {Jevremovi{\' c}}, D., {Short}, C.~I. et al. 2001, \aap, 369, L13

\bibitem[{{Doyle} \& {Mathioudakis}(1990)}]{Doyle90}
{Doyle}, J.~G. \& {Mathioudakis}, M. 1990, \aap, 227, 130

\bibitem[{{Doyle} {et~al.}(1992){Doyle}, {van der Oord}, \&
  {Kellett}}]{Doyle92}
{Doyle}, J.~G., {van der Oord}, G.~H.~J., \& {Kellett}, B.~J. 1992, \aap, 262,
  533

\bibitem[{{Drake}(1980)}]{Drake80}
{Drake}, S.~A. 1980, Ph.D.~Thesis, 72

\bibitem[{{Drake} \& {Ulrich}(1980)}]{DrakeUlrich80}
{Drake}, S.~A. \& {Ulrich}, R.~K. 1980, \apjs, 42, 351

\bibitem[{{Foing} {et~al.}(1994){Foing}, {Char}, {Ayres}, {Catala}, {Neff},
  {Zhai}, {Catalano}, {Cutispoto}, {Jankov}, {Rodono}, {Simon}, {Akan},
  {Aslanov}, {Avellar}, {Baudrand}, {Beust}, {Cao}, {Chatzichristou}, {Cuby},
  {Czarny}, {de La Reza}, {Dreux}, {Felenbok}, {Ferlet}, {Frasca}, {Floquet},
  {Ghosh}, {Guo}, {Guerin}, {Hao}, {Houdebine}, {Huang}, {Hubert}, {Hubert},
  {Huovelin}, {Hron}, {Ibanoglu}, {Jiang}, {Keskin}, {Lagrange-Henri},
  {Lecontel}, {Li}, {Mavridis}, {Nolthenius}, {Petrov}, {Savanov},
  {Scherbakov}, {Tuominen}, {Vidal-Madjar}, {Zhang}, \& {Zhang}}]{Foing94}
{Foing}, B.~H., {Char}, S., {Ayres}, T. et al. 1994, \aap, 292, 543

\bibitem[{{Garcia Alvarez}(2000)}]{Garcia-Alvarez00}
{Garcia Alvarez}, D. 2000, Irish Astronomical Journal, 27, 117

\bibitem[{{Gershberg}(1989)}]{Gershberg89}
{Gershberg}, R.~E. 1989, Memorie della Societa Astronomica Italiana, 60, 263

\bibitem[{{Giampapa}(1983)}]{Giampapa83}
{Giampapa}, M.~S. 1983, in ASSL Vol. 102: IAU Colloq. 71: Activity in Red-Dwarf
  Stars, 223--233

\bibitem[{{Gieske} \& {Griem}(1969)}]{Gieske69}
{Gieske}, H.~A. \& {Griem}, H.~R. 1969, \apj, 157, 963

\bibitem[{{Haisch} {et~al.}(1991){Haisch}, {Strong}, \& {Rodono}}]{Haisch91}
{Haisch}, B., {Strong}, K.~T., \& {Rodono}, M. 1991, \araa, 29, 275

\bibitem[{{Hawley} \& {Fisher}(1992)}]{Hawley92}
{Hawley}, S.~L. \& {Fisher}, G.~H. 1992, \apjs, 78, 565

\bibitem[{{Hawley} \& {Pettersen}(1991)}]{Hawley91}
{Hawley}, S.~L. \& {Pettersen}, B.~R. 1991, \apj, 378, 725

\bibitem[{{Houdebine}(1992)}]{Houdebine92}
{Houdebine}, E.~R. 1992, Irish Astronomical Journal, 20, 213

\bibitem[{{Houdebine} {et~al.}(1996){Houdebine}, {Mathioudakis}, {Doyle}, \&
  {Foing}}]{Houdebine96}
{Houdebine}, E.~R., {Mathioudakis}, M., {Doyle}, J.~G., \& {Foing}, B.~H. 1996,
  \aap, 305, 209

\bibitem[{{Jevremovi{\' c}}(1999)}]{Jevremovic99}
{Jevremovi{\' c}}, D. 1999, Ph.D.~Thesis, 187

\bibitem[{{Jevremovi{\' c}} {et~al.}(1998){Jevremovi{\' c}}, {Butler}, {Drake},
  {O'Donoghue}, \& {van Wyk}}]{Jevremovic98}
{Jevremovi{\' c}}, D., {Butler}, C.~J., {Drake}, S.~A., {O'Donoghue}, D., \&
  {van Wyk}, F. 1998, \aap, 338, 1057

\bibitem[{{Johns-Krull} {et~al.}(1997){Johns-Krull}, {Hawley}, {Basri}, \&
  {Valenti}}]{Johns-Krull97}
{Johns-Krull}, C.~M., {Hawley}, S.~L., {Basri}, G., \& {Valenti}, J.~A. 1997,
  \apjs, 112, 221

\bibitem[{{Joy} \& {Wilson}(1949)}]{Joy49}
{Joy}, A.~H. \& {Wilson}, R.~E. 1949, \apj, 109, 231

\bibitem[{{Katsova}(1990)}]{Katsova90}
{Katsova}, M.~M. 1990, Soviet Astronomy, 34, 614

\bibitem[{{Katsova} {et~al.}(1991){Katsova}, {Livshits}, {Butler}, \&
  {Doyle}}]{Katsova91}
{Katsova}, M.~M., {Livshits}, M.~A., {Butler}, C.~J., \& {Doyle}, J.~G. 1991,
  \mnras, 250, 402

\bibitem[{{Kundu} {et~al.}(1987){Kundu}, {Jackson}, {White}, \&
  {Melozzi}}]{Kundu87}
{Kundu}, M.~R., {Jackson}, P.~D., {White}, S.~M., \& {Melozzi}, M. 1987, \apj,
  312, 822

\bibitem[{{Kunkel}(1970)}]{Kunkel70}
{Kunkel}, W.~E. 1970, \apj, 161, 503

\bibitem[{{Kunkel}(1973)}]{Kunkel73}
---. 1973, \apjs, 25, 1

\bibitem[{{Linsky} {et~al.}(1982){Linsky}, {Bornmann}, {Carpenter}, {Hege},
  {Wing}, {Giampapa}, \& {Worden}}]{Linsky82}
{Linsky}, J.~L., {Bornmann}, P.~L., {Carpenter}, K.~G. et al. 1982, \apj, 260, 670

\bibitem[{{Lites} {et~al.}(1986){Lites}, {Neidig}, \& {Trujillo
  Bueno}}]{Lites86}
{Lites}, B.~W., {Neidig}, D.~F., \& {Trujillo Bueno}, J. 1986, in The Lower
  Atmosphere of Solar Flares; Proceedings of the Solar Maximum Mission
  Symposium, 101--116

\bibitem[{{Mirzoyan}(1984)}]{Mirzoyan84}
{Mirzoyan}, L.~V. 1984, Vistas in Astronomy, 27, 77

\bibitem[{{Moffett} \& {Bopp}(1976)}]{Moffet76}
{Moffett}, T.~J. \& {Bopp}, B.~W. 1976, \apjs, 31, 61

\bibitem[{{Montes} {et~al.}(1999){Montes}, {Saar}, {Collier Cameron}, \&
  {Unruh}}]{Montes99}
{Montes}, D., {Saar}, S.~H., {Collier Cameron}, A., \& {Unruh}, Y.~C. 1999,
  \mnras, 305, 45

\bibitem[{{Nelson} {et~al.}(1986){Nelson}, {Robinson}, {Slee}, {Ashley},
  {Hyland}, {Tuohy}, {Nikoloff}, \& {Vaughan}}]{Nelson86}
{Nelson}, G.~J., {Robinson}, R.~D., {Slee}, O.~B. et al. 1986,
  \mnras, 220, 91

\bibitem[{{Pallavicini} {et~al.}(1990){Pallavicini}, {Tagliaferri}, \&
  {Stella}}]{Pallavicini90}
{Pallavicini}, R., {Tagliaferri}, G., \& {Stella}, L. 1990, \aap, 228, 403

\bibitem[{{Pettersen}(1989)}]{Pettersen89}
{Pettersen}, B.~R. 1989, \solphys, 121, 299

\bibitem[{{Redman} \& {Suemoto}(1954)}]{Redman54}
{Redman}, R.~O. \& {Suemoto}, Z. 1954, \mnras, 114, 524

\bibitem[{{Svestka}(1963)}]{Svestka63}
{Svestka}, Z. 1963, Bulletin of the Astronomical Institutes of Czechoslovakia,
  14, 234

\bibitem[{{Tandberg-Hanssen} \& {Emslie}(1988)}]{Tandberg-Hanssen88}
{Tandberg-Hanssen}, E. \& {Emslie}, A.~G. 1988, {The physics of solar flares}
  (Cambridge and New York, Cambridge University Press, 1988, 286 p.)

\bibitem[{{Torczon}(1991)}]{Torczon91}
{Torczon}, V. 1991, SIAM J.Optimization, 1, 123

\bibitem[{{Torczon}(1992)}]{Torczon92}
---. 1992, {} (Tech. Report 92-9, Dep. of Mathematical Sciences, Rice
  University, Houston)

\bibitem[{{{\v S}vestka}(1972)}]{Svestka72}
{{\v S}vestka}, Z.~.~K. 1972, \araa, 10, 1

\bibitem[{{Vidal}(1964)}]{Vidal64}
{Vidal}, C.~R. 1964, Zeitschrift Naturforschung Teil A, 19, 947

\bibitem[{{Wilson}(1978)}]{Wilson78}
{Wilson}, O.~C. 1978, \apj, 226, 379

\bibitem[{{Worden}(1983)}]{Worden83}
{Worden}, S.~P. 1983, in ASSL Vol. 102: IAU Colloq. 71: Activity in Red-Dwarf
  Stars, 207--220

\bibitem[{{Wuelser} \& {Marti}(1989)}]{Wuelser89}
{Wuelser}, J. \& {Marti}, H. 1989, \apj, 341, 1088

\bibitem[{{Zarro} \& {Canfield}(1989)}]{Zarro89}
{Zarro}, D.~M. \& {Canfield}, R.~C. 1989, \apjl, 338, L33

\bibitem[{{Zarro} \& {Zirin}(1985)}]{Zarro85}
{Zarro}, D.~M. \& {Zirin}, H. 1985, \aap, 148, 240

\end{thebibliography}

\end{document}